# Analytical Solution of Brillouin Amplifier Equations for lossless medium

Fikri Serdar Gökhan[a],* Hasan Göktaş,[b]
[a]Department of Electrical and Electronic Engineering, Alanya Alaaddin Keykubat University, Kestel, Alanya, Antalya, Turkey, [b]Department of Electrical and Computer Engineering, [b]Department of Electrical and Electronic Engineering, Harran University, Sanliurfa, 6300, Turkey

**ABSTRACT**

In order to explain pump depletion in Stimulated Brillouin scattering (SBS), coupled intensity equations describing the interaction of pump and stokes waves in Brillouin medium, must be solved simultaneously. Since this problem has well-defined boundary conditions, such a mathematical problem is known as the two-point boundary value problem. Conventional solution techniques leads transcendental equation which results implicit solution. In this paper, we accurately define Pump and Stokes evolution in lossless medium in terms of conserved quantity and proposed the solution of this conserved quantity using the asymptotic theory. Regarding with the saturation region, the gain approximation of Brillouin Fiber Amplifier (BFA) for the lossless medium, is introduced for the first time to our best of knowledge.

## 1. INTRODUCTION

Stimulated Brillouin scattering (SBS) is the most efficient nonlinear amplification mechanism in optical fibers, in which a large gain may be obtained under the pump power of several milliwatts[1]. This has led to the design of (BFA) and has been implemented in a wide range of applications, such as an active filter due to its narrowband amplification feature[2] or in the control of pulse propagation in optical fibers[3]. The BFA can also be used to measure strain and temperature[4] which has led to the design of distributed Brillouin sensors (DBS). In these type of sensors, strain and temperature can be measured along the whole fiber length[5]. In the BFA configuration, SBS can be used for efficient narrow band amplification when the Stokes wave is input from the rear (opposite to the pump) end of the fiber. Interaction between the pump and the Stokes wave due to SBS is described by a system of ordinary differential equations (ODEs)[6]. The system of ODEs for BFAs has well-defined boundary conditions: $P_p(0) = P_0$ and $P_S(L) = P_{Stokes}$. Such a mathematical problem is known as the two-point boundary value problem. Since the boundary value of the $P_p(L)$ and/or $P_s(0)$, with $L$ being the fiber length, is undetermined, such systems of non-linear ODEs is typically addressed numerically. The exact analytical solution to the system of ODEs is known only for lossless media[7], with the exception of an analytical solution of integration constant $C$. However, obtained expression is a transcendental equation giving the unknown quantity $P_S(0)/P_P(0)$ in terms of the known quantities $P_P(0)$ and $P_S(L)$. In this paper, we focus on the derivation of the pump and stokes evolution equations depending on the conserved quantity $C$ and proposed the solution of this conserved quantity using asymptotic theory[8]. The solution of conserved quantity is used for the derivation of analytical solution of Brillouin Amplifier Equations in a lossy medium.

## 2. THEORETICAL MODEL

The coupled ODEs for the evolution of the intensities of pump $I_P$ and Stokes $I_S$ in a lossless medium can be written as,

$$dI_p / dz = -g_B I_p I_s$$
$$dI_s / dz = -g_B I_p I_s \quad (1)$$

here $0 \leq z \leq L$ is the propagation distance along the optical fiber of the total length $L$, $g_B$ is the Brillouin gain coefficient. Note that, here we assume a Stokes wave launched from the rear end of the fiber. Then the known values of the input pump intensity $I_P(0) = I_{P0}$ and the input Stokes intensity $I_S(L) = I_{SL}$ are the boundary values. The geometry of an SBS amplifier is shown in figure 1.

*serdar.gokhan@alanya.edu.tr; phone +90 (242) 5106120-2560; fax: +90 (242) 5106009; alanya.edu.tr

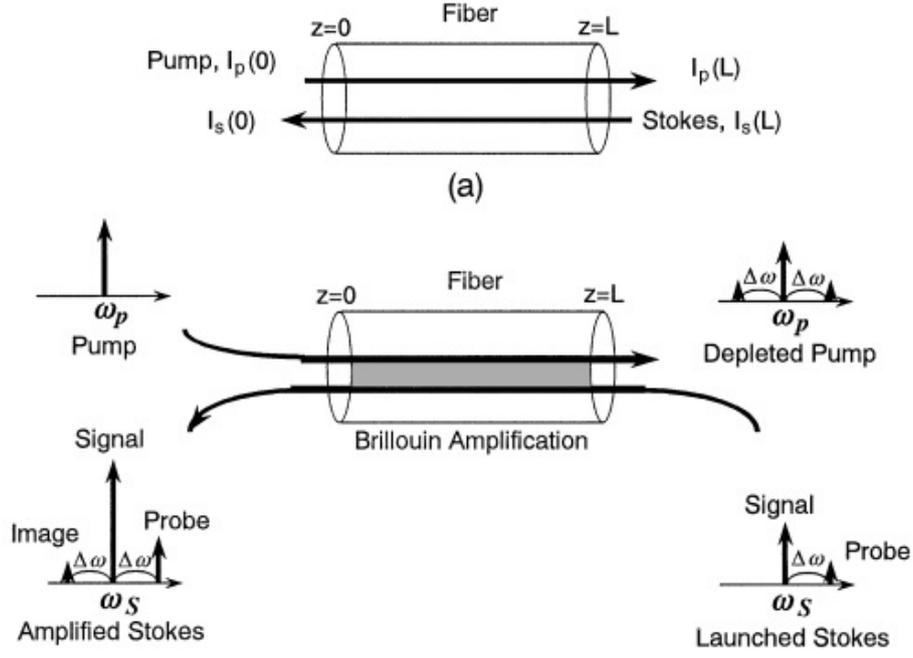

Figure 1. Geometry of an SBS amplifier.

The solution of (1) is derived in Ref. [9], which leads to the solution of,

$$I_p(z) = cI_{p0} \cdot \left[ I_{p0} + (c - I_{p0})\exp(-cg_B z) \right]^{-1}$$

$$I_s(z) = c(I_{p0} - c_1) \cdot \left[ I_{p0} \exp(cg_B z) + (c - I_{p0}) \right]^{-1}$$

(2)

where $I_p(z) = I_s(z) + I_{p0} - I_{s0}$ and $c = I_p(z) - I_s(z) = I_{p0} - I_{s0}$ is the conserved quantity. (3)

To explicitly find the value of parameter $c$ we approximately solve the equation:

$$I_{sL} = c(I_{p0} - c) \cdot \left[ I_{p0} \exp(cg_B L) + (c - I_{p0}) \right]^{-1}$$

(4)

using boundary condition $I_{sL}$. Within the paper, we will show two different solutions to the $c$ depending on the High Gain Region ($c_1$) and Saturation region ($c_2$).

## 3. SOLUTION

### 3.1 Solution of $c$ for High-Gain Region

In this region, let the solution of $c = c_1$,

$$I_{sL} = \frac{c_1(I_{p0} - c_1)}{\left[ I_{p0} \exp(c_1 g_B L) + (c_1 - I_{p0}) \right]}$$

(5)

$$\frac{I_{sL}}{I_{p0}} = \frac{\frac{c_1}{I_{p0}}\left(1 - \frac{c_1}{I_{p0}}\right)}{\left[ \exp\left(\frac{c_1}{I_{p0}} g_B I_{p0} L\right) + \left(\frac{c_1}{I_{p0}} - 1\right) \right]}$$

(6)

Let $\varepsilon = I_{sL}/I_{p0}$, $\kappa = g_B I_{p0} L$, $c_0 = c_1/I_{p0}$,

$$\varepsilon = \frac{c_0(1-c_0)}{e^{c_0\kappa} - (1-c_0)} \tag{7}$$

Denote $y = c_0 \kappa$ and rewrite (7) as

$$\frac{e^y - \left(1 - \frac{y}{\kappa}\right)}{y\left(1 - \frac{y}{\kappa}\right)} = \frac{1}{\varepsilon\kappa} \tag{8}$$

Using the definition $\ln\left(\frac{1}{\varepsilon\kappa}\right) = \Lambda$, we rewrite (8) as

$$\ln\left(e^y - \left(1 - \frac{y}{\kappa}\right)\right) - \ln y - \ln\left(1 - \frac{y}{\kappa}\right) = \Lambda \tag{10}$$

$$\ln\left(e^y\left(1 - \frac{\left(1 - \frac{y}{\kappa}\right)}{e^y}\right)\right) - \ln y - \ln\left(1 - \frac{y}{\kappa}\right) = \Lambda \tag{11}$$

$$y + \ln\left(\left(1 - \frac{1}{e^y}\right)\left(1 + \frac{y}{\kappa(e^y - 1)}\right)\right) - \ln y - \ln\left(1 - \frac{y}{\kappa}\right) = \Lambda \tag{12}$$

$$y + \ln\left(1 - \frac{1}{e^y}\right) + \ln\left(1 + \frac{y}{\kappa(e^y - 1)}\right) - \ln y - \ln\left(1 - \frac{y}{\kappa}\right) = \Lambda \tag{13}$$

We look for the solution of (13) in the form:

$$y = \Lambda(1 + \xi) \tag{14}$$

where $\xi \ll 1$. Next, substitute (14) in (13):

$$\Lambda + \Lambda\xi + \ln\left(1 - \frac{1}{e^\Lambda}\frac{1}{e^{\Lambda\xi}}\right) + \ln\left(1 + \frac{\Lambda + \Lambda\xi}{\kappa(e^{\Lambda+\Lambda\xi} - 1)}\right) - \ln\Lambda - \ln(1+\xi) - \ln\left(1 - \frac{\Lambda(1+\xi)}{\kappa}\right) = \Lambda \tag{15}$$

$$\Lambda + \Lambda\xi + \ln\left(1 - \frac{1}{e^\Lambda}\frac{1}{e^{\Lambda\xi}}\right) + \ln\left(1 + \frac{\Lambda + \Lambda\xi}{\kappa(e^{\Lambda+\Lambda\xi} - 1)}\right) - \ln\Lambda - \ln(1+\xi) - \ln\left(1 - \frac{\Lambda}{\kappa}\right) - \ln\left(1 - \frac{\Lambda\xi}{\kappa\left(1 - \frac{\Lambda}{\kappa}\right)}\right) = \Lambda \tag{16}$$

Following the asymptotic theory, we discard terms linear in $\xi$ (using series expansion of the ln function) and obtain,

$$\Lambda\xi + \ln\left(1 - \frac{1}{e^\Lambda}\right) + \ln\left(1 + \frac{\Lambda}{\kappa(e^\Lambda - 1)}\right) - \ln\Lambda - \ln\left(1 - \frac{\Lambda}{\kappa}\right) = 0 \quad \text{which results in,} \tag{17}$$

$$\xi \approx \frac{1}{\Lambda}\left(\ln\frac{\Lambda\left(1 - \frac{\Lambda}{\kappa}\right)}{\left(1 - \frac{1}{e^\Lambda}\right)\left(1 + \frac{\Lambda}{\kappa(e^\Lambda - 1)}\right)}\right) \qquad \text{Finally,} \tag{18}$$

$$y = c_0 \kappa = \Lambda + \Lambda \xi = \Lambda + \left( \ln \frac{\Lambda\left(1 - \frac{\Lambda}{\kappa}\right)}{\left(1 - \frac{1}{e^\Lambda}\right)\left(1 + \frac{\Lambda}{\kappa(e^\Lambda - 1)}\right)} \right) \qquad (19)$$

$$c_0 \approx \frac{1}{\kappa}\left( \Lambda + \left( \ln \frac{\Lambda\left(1 - \frac{\Lambda}{\kappa}\right)}{\left(1 - \frac{1}{e^\Lambda}\right)\left(1 + \frac{\Lambda}{\kappa(e^\Lambda - 1)}\right)} \right) \right) \qquad (20)$$

$$c_1 \approx \frac{1}{\kappa}\left( \Lambda + \ln\left(\Lambda\left(1 - \frac{\Lambda}{\kappa}\right)\right) - \ln\left(1 - \frac{1}{e^\Lambda}\right) - \ln\left(1 + \frac{\Lambda}{\kappa(e^\Lambda - 1)}\right) \right) I_{p0} \qquad (21)$$

where $\varepsilon = I_{sL}/I_{p0}$, $\kappa = g_B I_{p0} L$, $\Lambda = -\ln(\varepsilon\kappa)$.

In figure 2, the comparison of analytical solution of (20) with the real solution of (7) is plotted. As can be seen from the figure 2, when the $\varepsilon$ decreases, the accuracy of $c_0$ increases.

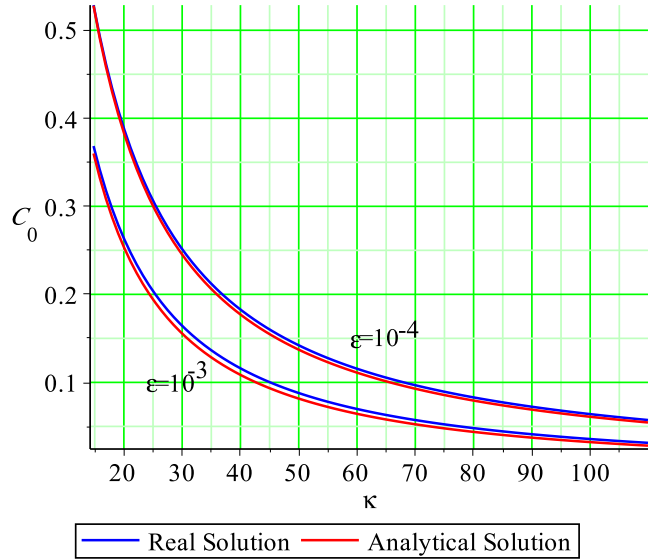

Figure 2. The comparison of (20) with the real solution of (7). $A_{eff} = 80\mu m^2$, $g_B = 1.091 \times 10^{-11}$.

## 3.2 Solution of c for Saturation Region

It can be shown that in saturation region $\Lambda < 0$ and $I_{p0} - c_2 \simeq I_{p0}$. The exchange of the powers show high nonlinearity. In this region, let the solution of $c = c_2$. In this case, the solution of (5) can be analytically defined as,

$$I_{sL} = \frac{c_2(I_{p0} - c_2)}{\left[I_{p0}\exp(c_2 g_B L) - (I_{p0} - c_2)\right]} = \frac{c_2}{\left[\frac{I_{p0}}{I_{p0}-c_2}\exp(c_2 g_B L) - 0.99\right]} \approx \frac{c_2}{\left[\exp(c_2 g_B L) - 0.99\right]} \quad (22)$$

The solution of (22) is[10]:

$$c_2 \simeq -0.99 I_{sL} - \frac{1}{g_B L} LambertW\left(-I_{sL} g_B L \cdot e^{-0.99 I_{sL} g_B L}\right) \quad (23)$$

where LambertW is a function which satisfies LambertW($x$) exp (Lambert W($x$) ) = $x$.

## 4. EXPERIMENTAL

In order to show the validity of the (21), we have used the modified experimental setup of Ref. [6]. In our experimental setup (Fig. 3), similar to Ref. [9], a tunable laser with $\lambda_p$=1549.5 nm was used together with an erbium-doped fiber amplifier (OA) to generate up to 80 mW of pump power. Stokes and pump sources are erbium-doped fiber lasers, whose linewidths are smaller than 100 kHz. Their frequency difference is controlled with a phase-locked loop and is locked to the Brillouin frequency, $\nu_B$, of the standard single-mode optical fiber under test. Power meters were used to monitor the input pump power $P_{P0}$, the transmitted pump power $P_{pL}$, the launched Stokes power $P_{sL}$, and the amplified Stokes power $P_{s0}$. To keep the attenuation low as possible, 2 km fiber length was experimentally studied.

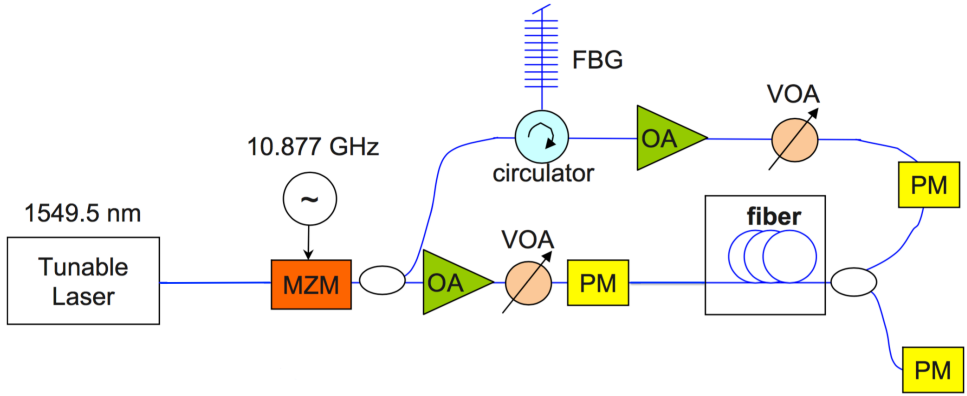

Figure 3. Experimental setup for BFA measurements: MZM, Mach–Zehnder modulator; PM, Powermeter; VOA, variable optical attenuator; OA(EDFA), optical amplifier; FBG, fiber Bragg grating used to select the upper modulation sideband as a Stokes signal.

The experimental results are plotted in figure 4 together with theoretical predictions. We find a high level of agreement for the transmitted pump for all input pump levels above critical pump power (Fig. 4(a)). As for the amplifier gain $G_{BFA}$, we find an excellent agreement between predictions from our analytical formula of gain computed with (21) and the measured gain (Fig. 4(b)). We note that (21) is applicable only when the pump power exceeds the critical value, $P_{p0} > P_{cr} \approx \left(\Lambda + \frac{A_{eff}}{0.00095 \times g_B L}\Lambda\right)$.

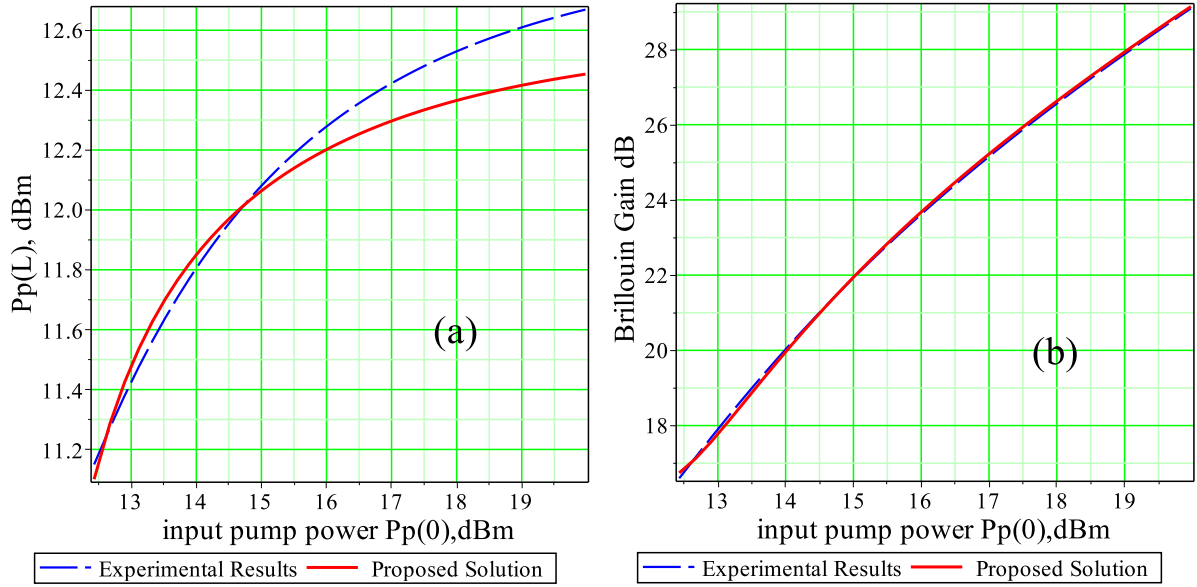

Figure 4. (a) Transmitted pump power $P_p(L)$ and (b) BFA gain $G_{BFA}$ versus the input pump power $Pp(0)$ in the High Gain Region. Dashed Blue, experimental data; thick red solid curves, predictions of the $P_p(L)$ using analytical formula (21). $L$=2 km, $P_{SL}$ =0.1 mW, $A_{eff}$ = 80μm$^2$, $g_B$=1.091x10$^{-11}$.

In figure 5, comparison of the (23) with the numerical solution using root finding algorithm[10] is plotted. It can be concluded from the figures 5a&5b, (23) is valid when Λ<0. In figure 6, BFA gain ($G_{BFA}$) versus the input pump power $P_p(0)$ in the High Gain Region is shown. From the figure 6, we find a high level of agreement between predictions from our analytical formula of gain computed with (23) and the measured gain. It can be concluded that for the same specific length, agreement between the gain of BFA with the measurement is better if the input pump power increases.

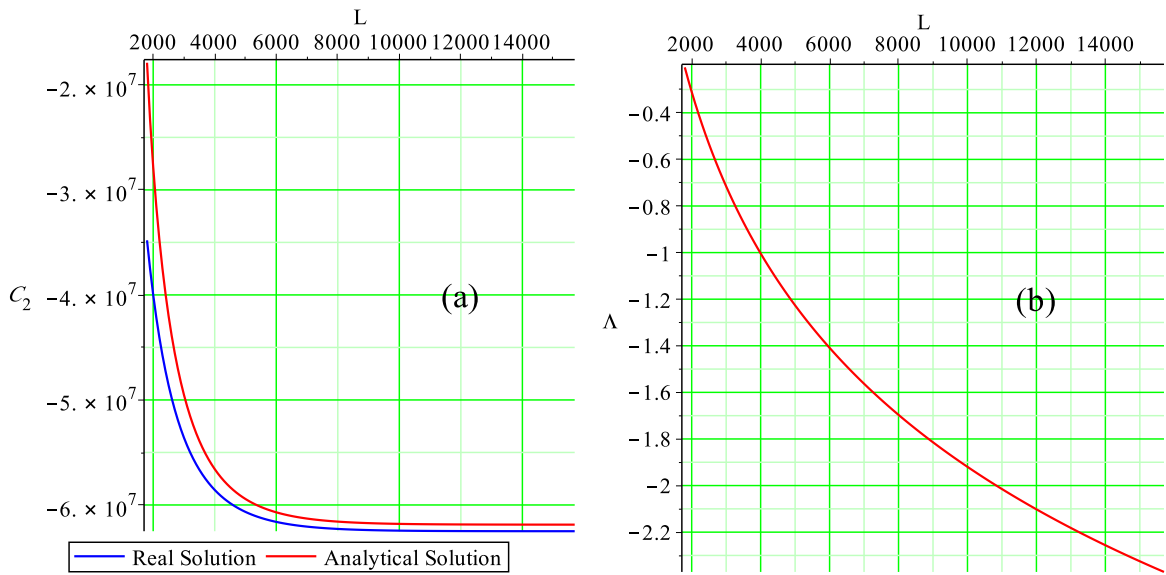

Figure 5. (a) The comparison of (23) with the real solution with using root finding algorithm. (b) The variation Λ with fiber length. $A_{eff}$ = 80μm$^2$, $g_B$=1.091x10$^{-11}$

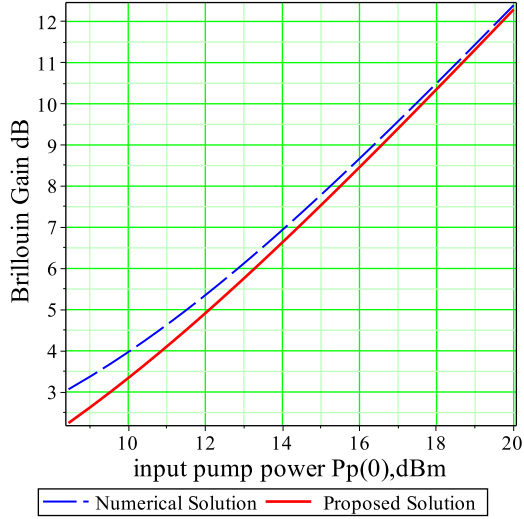

Figure 6. BFA gain $G_{BFA}$ versus the input pump power $P_p(0)$ in the High Gain Region. Dashed Blue, experimental data; thick red solid curves, predictions of the $P_p(L)$ using analytical formula (23). $L$=2 km km, $P_{SL}$=6 mW, $A_{eff}$ = 80μm$^2$, $g_B$=1.091x10$^{-11}$.

In Table 1, the usage of the equations described so far is briefly listed. The relation (21) is valid in the high-gain region where the pump power should be more than the $P_{cr}$. The relation (23) is valid in the saturation region where Λ<0.

Table 1, Brief usage of the Equations.

| Criteria | Region | Equation |
|---|---|---|
| $P_{p0} > P_{cr} \approx \left(\Lambda + \frac{A_{eff}}{0.00095 \times g_B L}\Lambda\right)$ | High-Gain Region | (21) |
| $\Lambda < 0$ | Saturation Region | (23) |

## 5. CONCLUSION

We have presented an approximate analytical solutions to the system of SBS equations in a lossless medium in two different regimes, namely, high gain and saturation regions. The limits of the three separation conditions are determined and the respective BFA gain to these regions are accurately defined. Especially, for the lossless medium, the gain approximation of BFA for the saturation region, is introduced for the first time to our best of knowledge. The results obtained can be practically used to optimize performance of Brillouin fiber amplifiers where the medium loss is negligible. And, the results can also be used in sensing applications which especially employs the high-gain region and fiber lengths less than 2km.